\documentclass[sigplan,nonacm]{acmart}

\usepackage{xcolor}
\usepackage[pdf]{graphviz}
\usepackage{subcaption}
\usepackage{listings, listings-juno, listings-rust, listings-scheduler}
\usepackage{pifont}
\usepackage{tikz}
\usepackage{tabularx}
\usetikzlibrary{patterns}
\newcommand{\cmark}{\textcolor{green}{\ding{51}}}%
\newcommand{\xmark}{\textcolor{red}{\ding{55}}}%

\definecolor{backred}{HTML}{FF8787}
\definecolor{backgreen}{HTML}{87FF87}
\definecolor{backblue}{HTML}{8787FF}

\definecolor{textred}{HTML}{e93f3f} 
\definecolor{textgreen}{HTML}{46D946} 
\definecolor{textblue}{HTML}{4646D9} 

\usepackage{pgfplots}
\pgfplotsset{compat=1.18}

\AtBeginDocument{%
  }

\setcopyright{acmlicensed}
\copyrightyear{2025}
\acmYear{2025}
\acmDOI{XXXXXXX.XXXXXXX}
\acmConference[ASPLOS '26]{}{}{}
\acmISBN{978-1-4503-XXXX-X/2018/06}

\begin{document}

\title{Hercules: A Compiler for Productive Programming of Heterogeneous Systems}

\author{Russel Arbore}
\email{rarbore2@illinois.edu}
\affiliation{
  \institution{University of Illinois Urbana-Champaign}
  \city{Urbana}
  \state{Illinois}
  \country{USA}
}
\author{Aaron Councilman}
\email{aaronjc4@illinois.edu}
\affiliation{
  \institution{University of Illinois Urbana-Champaign}
  \city{Urbana}
  \state{Illinois}
  \country{USA}
}
\author{Xavier Routh}
\email{xrouth2@illinois.edu}
\affiliation{
  \institution{University of Illinois Urbana-Champaign}
  \city{Urbana}
  \state{Illinois}
  \country{USA}
}
\author{Ryan Ziegler}
\email{ryanjz2@illinois.edu}
\affiliation{
  \institution{University of Illinois Urbana-Champaign}
  \city{Urbana}
  \state{Illinois}
  \country{USA}
}
\author{Praneet Rathi}
\email{prathi3@illinois.edu}
\affiliation{
  \institution{University of Illinois Urbana-Champaign}
  \city{Urbana}
  \state{Illinois}
  \country{USA}
}
\author{Vikram Adve}
\email{vadve@illinois.edu}
\affiliation{
  \institution{University of Illinois Urbana-Champaign}
  \city{Urbana}
  \state{Illinois}
  \country{USA}
}
\renewcommand{\shortauthors}{Arbore et al.}



\begin{abstract}
Modern computing systems increasingly rely on composing heterogeneous devices to improve performance and efficiency. Programming these systems is often unproductive: algorithm implementations must be coupled to system-specific logic, including device-specific optimizations, partitioning, and inter-device communication and synchronization, which requires developing different programs for different system configurations. We propose the Juno language, which represents general purpose applications in an imperative form that can be transformed into parallel, optimized, system-specific code using an expressive and granular imperative scheduling language. We also introduce the Hercules compiler, which uses a novel intermediate representation to represent general and device-specific parallel code in a manner that is easy to analyze and manipulate using schedules. Our system  achieves competitive performance with hand-optimized device-specific code (geomean speedups of $1.25\times$ and $1.48\times$ on the CPU and GPU) and significantly outperforms a prior general purpose heterogeneous programming system (geomean speedups of $9.31\times$ and $16.18\times$ on the CPU and GPU).
\end{abstract}

\maketitle

\section{Introduction}

Heterogeneous computing systems are becoming increasingly necessary to improve computing performance and efficiency. These systems exploit specialization and parallelism to out-perform other options. However, this additional architectural complexity induces programming system complexity. Low level programming interfaces, such as CUDA, OpenCL, and Vulkan, are the norm for accelerators~\cite{cuda_release, tsuchiyama2011opencl, vulkan_spec}. These systems require explicit management of low level details, such as data movement and synchronization, which are critical to achieve optimal performance. Manually tuning these details sacrifices portability, adds complexity, and requires hardware knowledge during application development. Although these programming interfaces are widely used, their use can hardly be called productive.

Several high level programming systems have been proposed to target generic programs onto heterogeneous systems. OpenMP~\cite{openmp} provides compiler directives and a runtime library to enable easy parallelization of C, C++, and Fortran applications, with version 4.0 adding support for targeting GPUs~\cite{openmp4}. OpenMP annotations can grow complex: programmers may need to identify which variables are private to each loop iteration, what reductions are present, what memory objects are accessed, and other properties in order to generate fast code. Thus, OpenMP still requires application programmers possess hardware knowledge at development time.

Delite \cite{delite} implements DSLs by compiling programs to an intermediate representation composed of parallel patterns. Delite provides many transformations, code generators, and other components available for development of DSLs, but existing Delite DSLs do not expose control over those components to their end users. This choice improves programmability, but sacrifices performance opportunities when the compiler makes an incorrect optimization decision. 

Braid \cite{braid} is a compiler that uses static staging to perform device partitioning. This approach enables safe manipulation of application code, but these manipulations occur inside the application language, coupling application logic to a fixed lowering strategy. Additionally, Braid does not provide an IR for optimizing device-agnostic parallel patterns, instead primarily using static staging to represent device specific constructs, such as graphics pipelines on the GPU.

HPVM \cite{hpvm2017, hpvm2022, hpvm2fpga.asap22} is a general purpose compiler targeting heterogeneous systems composed of CPUs, GPUs, and FPGAs. HPVM IR is a low level intermediate representation for parallel and heterogeneous code, but only represents parallel loops that compute non-overlapping results---it fails to represent parallel associative reductions, among other patterns. HPVM IR requires explicit annotations describing what loops are parallel, what memory objects are the inputs and outputs of each function, and what functions will be lowered to device kernels. These properties are neither automatically inferrable nor easily transformable. This makes many optimizations, such as loop fusion and fission, difficult to perform automatically. The most commonly used HPVM frontend, Hetero-C++, requires the programmer to manually add these annotations. 

None of these prior systems simultaneously provide a productive, target agnostic application language and a productive mechanism to control how an application is lowered onto a heterogeneous system.

Another approach to programming heterogeneous systems is to use domain specific languages (DSLs)~\cite{dsls, halide, tensorflow2015-whitepaper, pytorch, jax2018github}. DSLs make writing programs for a single domain easier for developers, and this specificity also makes it easier to implement compiler optimizations, which often leverage domain-specific information. However, many applications involve algorithms from multiple domains and different DSLs are not necessarily designed to work well together, hurting both programmability and the potential for global optimization.

To enable productive development of applications for heterogeneous systems, we believe a separation is necessary wherein the application can be developed without hardware knowledge, and therefore without hardware-specific restructuring or optimization, and the developer has control to later heavily optimize their application for a specific heterogeneous system. This separates the roles of the application and performance engineers and should enable productive workflows for both that do not interfere with each other.

To this end, we have developed a language called Juno with two components: an application language for expressing program logic and a scheduling language for optimizing Juno applications. Applications are easy to write, since Juno features standard imperative control flow and mutation. Juno does \textit{not} contain primitives for expressing parallelism or heterogeneity---Juno applications are completely hardware independent. Juno schedules are imperative and granular---schedules can refer to individual statements in a Juno program and apply transformations to those selections. Since schedules are de-coupled from application logic, a single application can have multiple schedules for different machine configurations. To the best of our knowledge, this is the first proposed scheduling language for optimizing general purpose, imperative applications in a heterogeneous context.

We also present Hercules, a compiler that automatically extracts parallelism and heterogeneity from sequential code. The Hercules intermediate representation (IR) is based off of the sea of nodes IR~\cite{seaofnodes, hotspot} and uses value semantics to represent programs. Hercules IR explicitly represents parallel loops using fork-joins, a novel construct which encodes both data parallel and reduction computations. We implement several optimizations for Hercules IR, including loop optimizations such as tiling and fusion; these optimizations can be applied in specific places via a Juno schedule. By transforming the structure of fork-joins in Hercules IR, the compiler can expose a hierarchy of parallelism and memory communication to enable high quality device code generation. An application is restructured for heterogeneous execution by schedule controlled function inlining and outlining, where functions transition from representing logical units of code into units of device code that are compiled separately. Hercules IR functions are then lowered either into device code targeting multi-core CPUs or GPUs, or into host code that orchestrates the execution of device functions. Using Hercules, \textit{we compile unmodified, unspecialized applications onto both multi-core CPUs and NVIDIA GPUs, and we achieve competitive performance with prior hand-written baselines and significantly outperform HPVM, a prior general purpose heterogeneous programming system}.

Overall, we make the following contributions:

\begin{enumerate}
    \item The first highly granular scheduling language that enables fine-grained selections of functions, or parts of functions, for transformation, allowing device-specific schedules to exploit parallelism and heterogeneity in an unmodified, generic application, irrespective of the original code structure (Section~\ref{frontend}).
    \item A sea-of-nodes based intermediate representation that simultaneously enables device-agnostic optimizations for converting sequential code into parallel and heterogeneous code, and high performance code generation targeting CPUs and GPUs (Sections~\ref{hercules_ir}~and~\ref{device_codegen}).
    \item An evaluation of the performance of applications compiled with Hercules against a prior heterogeneous programming system and hand-written baselines. Hercules achieves $1.26\times$ and $1.47\times$ geomean speedups against prior hand-written programs on the CPU and GPU, respectively, and achieves $9.31\times$ and $16.18\times$ geomean speedups over HPVM on the CPU and GPU, respectively (Section~\ref{eval}).
\end{enumerate}

\section{Writing and Scheduling Heterogeneous Programs}
\label{frontend}

To enable productive programming of heterogeneous systems, we require a \emph{hardware-agnostic} application language~\cite{hpvm2fpga.asap22}. Algorithms should be expressed simply, in a manner oblivious to hardware, and should be able to be optimized for varying heterogeneous systems without modification. A truly hardware-agnostic language thus has several constraints:
\begin{enumerate}
    \item No I/O operations or OS interaction: different devices have different I/O and OS interaction capabilities, meaning use of I/O or OS interaction couples a piece of code to a specific device.
    \item No pointers or references: pointers and references do not map well across varying devices with varying memory models and separate memory regions. This implies that arbitrary memory allocations cannot be supported.
    \item No recursion or mutual recursion: recursion requires a mechanism for tracking calls (such as a stack), and this may not be possible on all devices.
\end{enumerate}
Notably, (2) suggests that such a language requires a \emph{value} semantics, where all pieces of data, including large collections like arrays, are treated as values that are passed between operations.
We propose an application language which fits these requirements (Section~\ref{sec:juno}) along with a scheduling language which is used to optimize it (Section~\ref{sec:juno-scheduler}). The scheduling language is granular and expressive, enabling sophisticated optimization of programs.

\subsection{The Juno Application Language}
\label{sec:juno}

The Juno application language is designed to satisfy the above constraints.
It can be used to express the computation of an application and is invoked from user Rust code, using an automatically generated interface (Section~\ref{rust_interface}); user Rust code is responsible for setup and operations that require OS interaction.
To achieve a familiar syntax, similar to Rust's, Juno uses a mutable value semantics \cite{hylo}, where a variable, element of an array, or field of a product can be mutated by assigning it a new value.
Figure~\ref{fig:juno_matmul} shows a Juno program which performs matrix multiplication. On line~\ref{line:juno_matmul_entry}, the \texttt{matmul} function is annotated \texttt{\#[entry]}, signifying a Rust interface for it should be generated. Next, \emph{dynamic constants} \texttt{n}, \texttt{m}, and \texttt{l} are introduced.
Dynamic constants represent values unknown at compile time but which will be known immediately before invoking a Juno program.
Since Juno uses a mutable value semantics, the compiler can statically determine the number of required allocations along with their sizes in terms of dynamic constants; the Rust host interface uses this information to perform memory allocations upfront (possibly on multiple devices) before executing Juno code.
The function body is standard: a result array is created and implicitly initialized to zeroes (line~\ref{line:juno_matmul_res}) and then written into.
Each of the loops is {labeled}, by \texttt{@outer}, \texttt{@middle}, and \texttt{@inner}. These labels are used in the scheduling language to reference specific regions of code on which to perform optimizations.

\begin{figure}
    \centering
\begin{lstlisting}[style=juno]
#[entry]                                        |\label{line:juno_matmul_entry}|
fn matmul<n, m, l: usize>                       |\label{line:juno_matmul_dcs}|
(a: f32[n, m], b: f32[m, l]) -> f32[n, l] {     |\label{line:juno_matmul_args}|
  let res : f32[n, l];                          |\label{line:juno_matmul_res}|
  |\color{textred}@outer| for i in 0..n {                        |\label{line:juno_matmul_i}|
    |\color{textgreen}@middle| for j in 0..l {                     |\label{line:juno_matmul_j}|
      |\color{textblue}@inner| for k in 0..m {                    |\label{line:juno_matmul_k}|
        res[i, j] += a[i, k] * b[k, j];         |\label{line:juno_matmul_compute}|
      }                                         |\label{line:juno_matmul_k_end}|
    }                                           |\label{line:juno_matmul_j_end}|
  }                                             |\label{line:juno_matmul_i_end}|
}                                               |\label{line:juno_matmul_body_end}|
\end{lstlisting}
    \caption{A Juno program to compute matrix multiplication.}
    \label{fig:juno_matmul}
\end{figure}

\begin{figure}
    \centering
\begin{lstlisting}[style=juno-scheduler]
// Parallelize by computing output array as 16 blocks
let par = matmul@outer \ matmul@inner;                   |\label{line:matmul_schedule_par}|
fork-chunk![4](par);                                     |\label{line:matmul_schedule_tile_par}|
let (outer, inner, _) = fork-reshape[[0,2],[1],[3]](par);|\label{line:matmul_schedule_reshape_par}|
parallelize!(outer \ inner);                             |\label{line:matmul_schedule_parallelize}|
let body = outline(inner);                               |\label{line:matmul_schedule_outline}|
cpu(body);                                               |\label{line:matmul_schedule_cpu}|
// Tile for cache, assuming 64B cache lines
fork-tile![16](body);                                    |\label{line:matmul_schedule_tile_cache}|
let (outer, inner) = fork-reshape[[0,2,4,1,3],[5]](body);|\label{line:matmul_schedule_reshape_cache}|
\end{lstlisting}
    \caption{A schedule to tile and parallelize the matrix multiplication in Figure~\ref{fig:juno_matmul}.}
    \label{fig:juno_matmul_schedule}
\end{figure}

\subsection{The Juno Scheduling Language}
\label{sec:juno-scheduler}

The Juno scheduling language is an imperative language that specifies the optimizations to apply to a program. Optimization passes can be applied on the entire module (referred to by \texttt{*}), whole functions, or \emph{labels} within a particular function.
In Juno, labels can be applied to any statement and the label then refers to that statement and any sub-expressions or sub-statements.
For instance, the \texttt{@middle} label in Figure~\ref{fig:juno_matmul} applies to the body of the \texttt{j} loop as well as the initialization and loop condition of that loop (lines~\ref{line:juno_matmul_j} through~\ref{line:juno_matmul_j_end}).
That section of code is then referred to by \texttt{matmul@middle}, where we prepend the function name as the same label can appear in multiple functions.
Labels can also be composed using set operations: relevant here is that \texttt{x $\setminus$ y} means set difference.
For instance, \texttt{matmul@middle $\setminus$ matmul@inner} refers to just the initialization, update, and condition of the \texttt{j} loop (excluding its body).

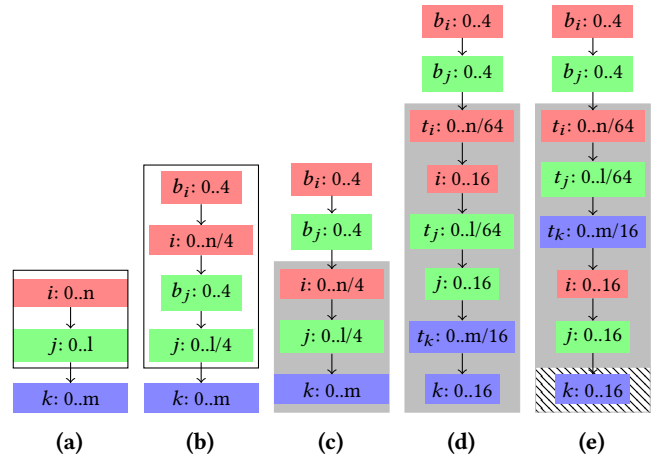
\begin{figure}
    \tikzset{node distance=.7cm}
    \centering\footnotesize
    \begin{subfigure}[b]{.18\linewidth}
        \centering
        \begin{tikzpicture}

            \node[rectangle, fill=backred, minimum width=\linewidth] (i) {$i$: 0..n};
            \node[rectangle, fill=backgreen, minimum width=\linewidth, below of=i] (j) {$j$: 0..l};
            \node[rectangle, fill=backblue, minimum width=\linewidth, below of=j] (k) {$k$: 0..m};
            \draw[draw=black] (-\linewidth / 2, -1cm) rectangle ++ (\linewidth, 1.3cm);
            \draw[->] (i) -- (j);
            \draw[->] (j) -- (k);
        \end{tikzpicture}
        \caption{}\label{fig:juno_matmul_loops_orig}
    \end{subfigure}
    \hfill
    \begin{subfigure}[b]{.18\linewidth}
        \centering
        \begin{tikzpicture}
            \draw[draw=black] (-\linewidth / 2, -2.4cm) rectangle ++ (\linewidth, 2.7cm);
            \node[rectangle, fill=backred, minimum width=.7\linewidth] (bi) {$b_i$: 0..4};
            \node[rectangle, fill=backred, minimum width=.9\linewidth, below of=bi] (i) {$i$: 0..n/4};
            \node[rectangle, fill=backgreen, minimum width=.7\linewidth, below of=i] (bj) {$b_j$: 0..4};
            \node[rectangle, fill=backgreen, minimum width=.9\linewidth, below of=bj] (j) {$j$: 0..l/4};
            \node[rectangle, fill=backblue, minimum width=\linewidth, below of=j] (k) {$k$: 0..m};
            \draw[->] (bi) -- (i);
            \draw[->] (i) -- (bj);
            \draw[->] (bj) -- (j);
            \draw[->] (j) -- (k);
        \end{tikzpicture}
        \caption{}\label{fig:juno_matmul_loops_chunk}
    \end{subfigure}
    \hfill
    \begin{subfigure}[b]{.18\linewidth}
        \centering
        \begin{tikzpicture}
            \draw[draw=lightgray, fill=lightgray] (.005, -.1cm) rectangle ++ (\linewidth, 2cm);
            \node[rectangle, fill=backblue, minimum width=\linewidth, anchor=south west] (k) at (0, 0) {$k$: 0..m};
            \node[rectangle, fill=backgreen, minimum width=.9\linewidth, above of=k] (j) {$j$: 0..l/4};
            \node[rectangle, fill=backred, minimum width=.9\linewidth, above of=j] (i) {$i$: 0..n/4};
            \node[rectangle, fill=backgreen, minimum width=.7\linewidth, above of=i] (bj) {$b_j$: 0..4};
            \node[rectangle, fill=backred, minimum width=.7\linewidth, above of=bj] (bi) {$b_i$: 0..4};
            \draw[->] (bi) -- (bj);
            \draw[->] (bj) -- (i);
            \draw[->] (i) -- (j);
            \draw[->] (j) -- (k);
        \end{tikzpicture}
        \caption{}\label{fig:juno_matmul_loops_chunk_reord}
    \end{subfigure}
    \hfill
    \begin{subfigure}[b]{.18\linewidth}
        \centering
        \begin{tikzpicture}
            \draw[draw=lightgray, fill=lightgray] (0, -.1cm) rectangle ++ (\linewidth, 4.1cm);
            \node[rectangle, fill=backblue, minimum width=.6\linewidth, anchor=south] (k) at (.5\textwidth, 0) {$k$: 0..16};
            \node[rectangle, fill=backblue, minimum width=.9\linewidth, above of=k] (tk) {$t_k$: 0..m/16};
            \node[rectangle, fill=backgreen, minimum width=.6\linewidth, above of=tk] (j) {$j$: 0..16};
            \node[rectangle, fill=backgreen, minimum width=.9\linewidth, above of=j] (tj) {$t_j$: 0..l/64};
            \node[rectangle, fill=backred, minimum width=.6\linewidth, above of=tj] (i) {$i$: 0..16};
            \node[rectangle, fill=backred, minimum width=.9\linewidth, above of=i] (ti) {$t_i$: 0..n/64};
            \node[rectangle, fill=backgreen, minimum width=.7\linewidth, above of=ti] (bj) {$b_j$: 0..4};
            \node[rectangle, fill=backred, minimum width=.7\linewidth, above of=bj] (bi) {$b_i$: 0..4};
            \draw[->] (bi) -- (bj);
            \draw[->] (bj) -- (ti);
            \draw[->] (ti) -- (i);
            \draw[->] (i) -- (tj);
            \draw[->] (tj) -- (j);
            \draw[->] (j) -- (tk);
            \draw[->] (tk) -- (k);
        \end{tikzpicture}
        \caption{}\label{fig:juno_matmul_loops_tile}
    \end{subfigure}
    \hfill
    \begin{subfigure}[b]{.18\linewidth}
        \centering
        \begin{tikzpicture}
            \draw[draw=lightgray, pattern=north west lines] (0, -.1cm) rectangle ++ (\linewidth, .6cm);
            \draw[draw=lightgray, fill=lightgray] (0, .5cm) rectangle ++ (\linewidth, 3.5cm);
            \node[rectangle, fill=backblue, minimum width=.6\linewidth, anchor=south] (k) at (.5\textwidth, 0) {$k$: 0..16};
            \node[rectangle, fill=backgreen, minimum width=.6\linewidth, above of=k] (j) {$j$: 0..16};
            \node[rectangle, fill=backred, minimum width=.6\linewidth, above of=j] (i) {$i$: 0..16};
            \node[rectangle, fill=backblue, minimum width=.9\linewidth, above of=i] (tk) {$t_k$: 0..m/16};
            \node[rectangle, fill=backgreen, minimum width=.9\linewidth, above of=tk] (tj) {$t_j$: 0..l/64};
            \node[rectangle, fill=backred, minimum width=.9\linewidth, above of=tj] (ti) {$t_i$: 0..n/64};
            \node[rectangle, fill=backgreen, minimum width=.7\linewidth, above of=ti] (bj) {$b_j$: 0..4};
            \node[rectangle, fill=backred, minimum width=.7\linewidth, above of=bj] (bi) {$b_i$: 0..4};
            \draw[->] (bi) -- (bj);
            \draw[->] (bj) -- (ti);
            \draw[->] (ti) -- (tj);
            \draw[->] (tj) -- (tk);
            \draw[->] (tk) -- (i);
            \draw[->] (i) -- (j);
            \draw[->] (j) -- (k);
        \end{tikzpicture}
        \caption{}\label{fig:juno_matmul_loops_tile_reord}
    \end{subfigure}
    \caption{The loop structure of \texttt{matmul} as it is scheduled. \textcolor{textblue}{Blue} blocks refer to the \textcolor{textblue}{\texttt{@inner}} loop, \textcolor{textgreen}{green} blocks refer to the \textcolor{textgreen}{\texttt{@middle}} loop, and \textcolor{textred}{red} blocks refer to the \textcolor{textred}{\texttt{@outer}} loop.}
    \label{fig:juno_matmul_loops}
\end{figure}

Figure~\ref{fig:juno_matmul_schedule} shows part of a schedule for optimizing the \texttt{matmul} function for CPUs.
Figure~\ref{fig:juno_matmul_loops} shows how the loop nest changes as we execute the schedule, with Figure~\ref{fig:juno_matmul_loops_orig} showing the initial three loops from the source Juno code.
The schedule operates as follows:
\begin{enumerate}
    \item First, we select only the outer two loops, shown with the black outline in the figure (line~\ref{line:matmul_schedule_par}), and we divide both of these loops into four chunks each (line~\ref{line:matmul_schedule_tile_par}). When the new dimensions are added above the old ones in the iteration space, we call it ``chunking.'' The resulting loop nest is shown in Figure~\ref{fig:juno_matmul_loops_chunk}.
    \item We then reorder the, now four, outer loops using the \texttt{fork-reshape} pass, so that the loops over blocks surround the loops within blocks.
    The arguments to this pass specify the loop indices (numbering with 0 as the top loop) in the order they should be transformed to; the result is shown in Figure~\ref{fig:juno_matmul_loops_chunk_reord}.
    After the reshape, the variable \texttt{outer} now refers to the entire loop nest and \texttt{inner} refers to just the inner three loops (with the gray shading behind them in the figure).
    \item Next, we mark the outer two loops (over the blocks) to be run in parallel as part of the host orchestration code (line~\ref{line:matmul_schedule_parallelize}) and then extract the inner loops into a device function that is assigned to the CPU (lines~\ref{line:matmul_schedule_outline} and~\ref{line:matmul_schedule_cpu}).
    \item Finally, we tile the inner three loops by 16, resulting in the loop nest shown in Figure~\ref{fig:juno_matmul_loops_tile} (line~\ref{line:matmul_schedule_tile_cache}). When the new dimensions are added below the old ones in the iteration space, we call it "tiling". We reorder the loops so that the loops over blocks surround the loops within blocks as shown in Figure~\ref{fig:juno_matmul_loops_tile_reord} (line~\ref{line:matmul_schedule_reshape_cache}).
    The variable \texttt{inner} is redefined to refer just to the final innermost loop (shown with the hatched background) so that it can be further optimized later (not shown).
\end{enumerate}

We highlight an important aspect of the Juno scheduling language that enables sophisticated schedules: passes can return values referring to specific regions of code. These region values are sets of nodes in the Hercules intermediate representation (Section~\ref{hercules_ir}).
Initially, every label in the application is defined as a region corresponding to the set of IR nodes generated by the Juno frontend corresponding to the labeled statements in the application code.

\section{The Hercules Intermediate Representation}
\label{hercules_ir}

We identify three design requirements for an IR that allows for comprehensive representation and transformation of parallel code targeting heterogeneous systems:
\begin{enumerate}
    \item It should enable easily working with common collection types, such as arrays and structures, by enabling optimization of code containing collections without worrying about device restrictions or conventions.
    \item It should provide an explicit representation of generic parallel code, as most heterogeneous architectures execute parallel code patterns.
    \item It should be highly analyzable and transformable, enabling fine-grained configuration of optimizations which allows users of Hercules to write schedules that directly transform their programs.
\end{enumerate}
To satisfy these requirements, we introduce a sea of nodes~\cite{seaofnodes} based intermediate representation, which, unlike prior work, represents collections as SSA values. We also introduce an explicit parallel loop construct called fork-joins. We use node labels to refer to specific parts of a program, allowing users to direct transformations with high granularity and specificity. To enable more aggressive optimization, we introduce node attributes, which after being inferred or manually applied give the compiler more information about a particular node.

\subsection{The Sea of Nodes and Value Semantics}
\label{seaofnodes}

\begin{figure}
    \centering
    \includegraphics[width=0.8\linewidth]{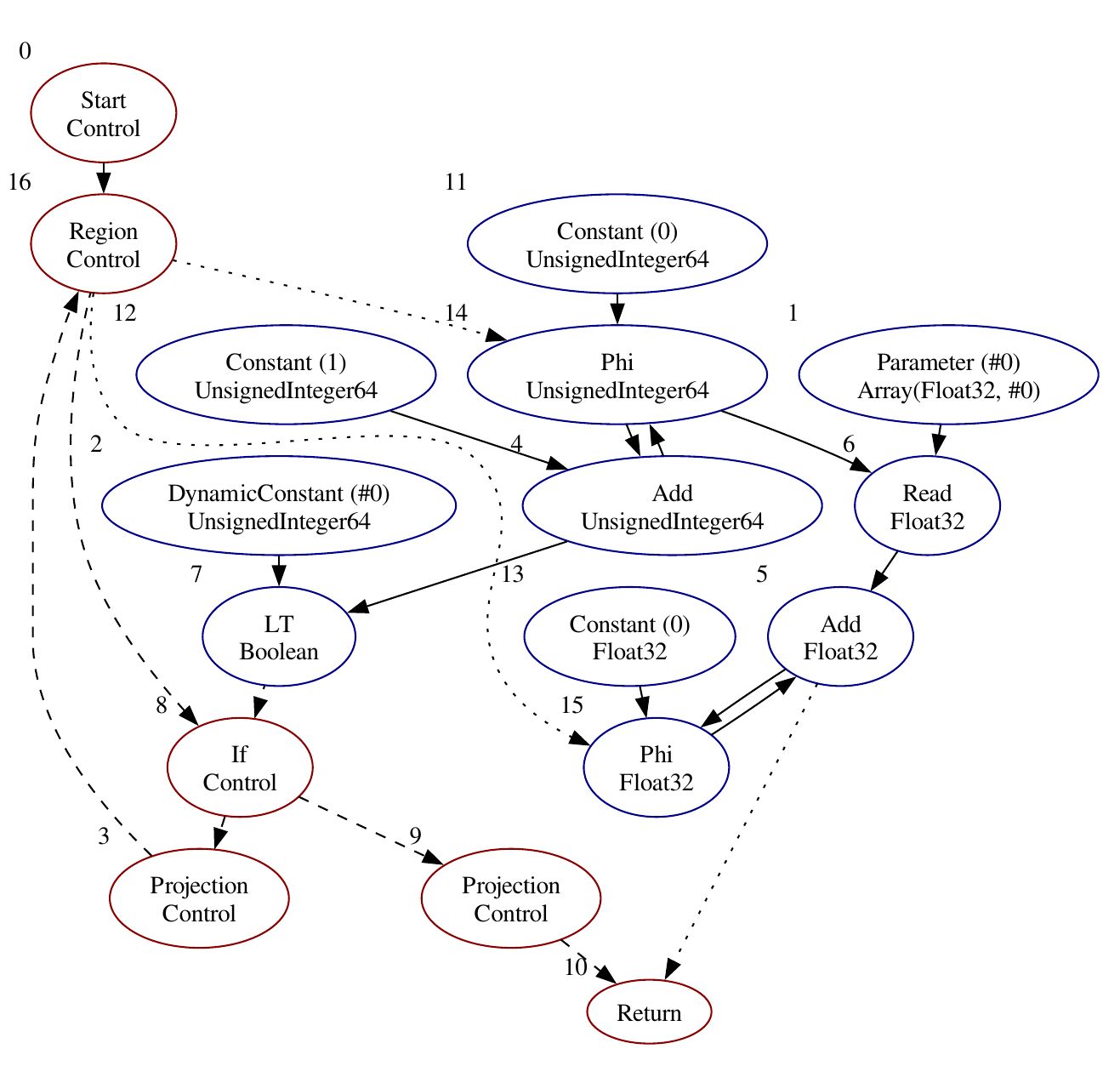}
    \caption{A Hercules IR function that adds all the elements in an array. Control nodes are red and data nodes are blue. Data dependencies are solid, control flow edges are dashed, control-data interactions are dotted.}
    \label{fig:arr_sum_ir}
\end{figure}

Hercules IR is based off of the sea of nodes~\cite{seaofnodes}. Every function is a single flow graph that mixes control and data nodes. Hercules IR is a single static assignment (SSA) IR: region nodes are analogous to the top of basic blocks, and in this analogy phi nodes use a region node to indicate what basic block they occupy. However, Hercules IR does not contain basic blocks---the interaction between data and control is sparse, as most data nodes do not have an explicit location. Semantically, a ``control token'' traverses the control flow subgraph, starting at the start node, to drive execution. Figure~\ref{fig:arr_sum_ir} shows the IR graph for a function that sums the elements of an array.

Hercules IR is based on immutable value semantics (commonly known as value semantics~\cite{hylo}), meaning that there are no first class references---all computations operate directly on data values and mutation is disallowed.
Collections are kept in SSA form, which is atypical; sea-of-nodes based IRs typically model memory as a single object that is shuffled between SSA variables, while in Hercules IR individual collections are modeled as separate values and data dependencies are explicit dataflow edges, which simplifies analysis and transformation. Hercules IR is not linear or affine---values can be cloned arbitrarily. Section~\ref{gcm} describes how we eliminate almost all clones of collections in practice.

\subsection{Collections and Dynamic Constants}
\label{dynconsts}

Hercules IR has three collection types: products (structures), summations (tagged unions), and arrays. Collection values are most commonly operated on by read and write nodes. Read nodes take in a collection value and a list of indices describing what element of the collection to read and output the value at that location. Similarly, a write node takes in a collection, a list of indices, and a data value, and outputs a new collection value that is the input collection modified by inserting the data input at the described location. Products are indexed by a static field index, summations by a static variant index (for accessing a variant of the summation type), and $N$ dimensional arrays by $N$ dynamic positions. Since indices are a logical representation corresponding to a collection type, Hercules is free to perform comprehensive analysis and optimization of code involving collections, such as scalar replacement of aggregates (SROA).

Many applications require arrays with dynamic size, but on heterogeneous systems it is uncommon to dynamically allocate memory on device. Rather, the host usually pre-allocates a dynamic amount of memory before running a device kernel. This dynamic size must be known before the device kernel is launched; we call such numbers \emph{dynamic constants}. The compiler does not know their value, but they do not vary at runtime since they are computed before device code is run. Dynamic constants are an explicit construct in Hercules IR: a function takes a number of dynamic constants as parameters, the type of an array includes its dimensions as dynamic constants, and the value of a dynamic constant can be used in computation via a dynamic constant node. These uses can be seen in Figure~\ref{fig:arr_sum_ir}: node 2 takes the first dynamic constant parameter of the function and produces its value and the function's parameter (node 1) has type \texttt{Array(Float32, \#0)}, which is an array of floats whose length is the first dynamic constant parameter.

\subsection{Representing Parallelism with Fork-Joins}
\label{forkjoins}

\begin{figure}
    \centering
    \includegraphics[width=0.8\linewidth]{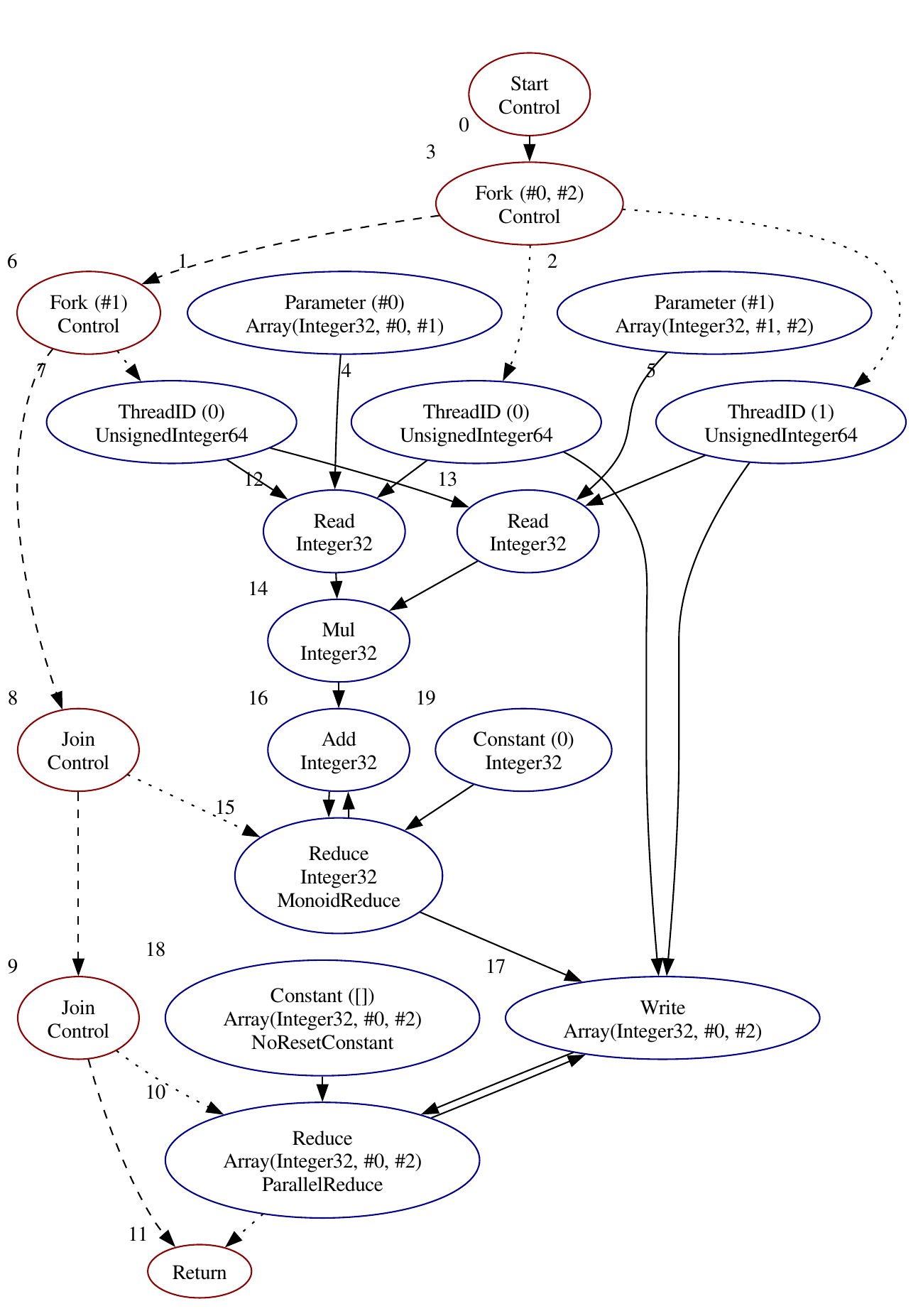}
    \caption{A Hercules IR function performing matrix multiplication using fork-joins.}
    \label{fig:matmul_ir}
\end{figure}

To enable effective reasoning about parallelism and reductions, we introduce \emph{fork-joins}, which represent SIMT code and generic (sometimes parallel) reductions. A fork-join consists of four kinds of nodes: a \emph{fork} node, a \emph{join} node, \emph{thread ID} nodes, and \emph{reduce} nodes. A fork node has a list of dynamic constant \emph{factors} which describe an $n$-dimensional grid of control tokens that are spawned when control reaches the fork. After all spawned tokens reach the join node, a single token exits the join. Thread ID nodes depend on fork nodes and produce a unique index, starting from zero, per spawned token along a single fork dimension. Finally, reduce nodes represent arbitrary reductions over the grid of tokens, taking an initial input and a reduction input. The reduction input describes how the partial reduction value and per-thread values are combined to compute the next step of the reduction. Semantically, the reduction occurs in increasing order of thread ID and factor dimension. Data computations not in reduction cycles can safely occur in parallel across threads. Figure~\ref{fig:matmul_ir} shows the IR for a matrix multiply routine that uses fork-joins to represent both the outer loop over output elements (node 3) and the inner dot product loop (node 6).

Fork-joins are a general loop representation, and many loops in real world code can be automatically transformed into fork-joins.
Hercules includes a pass, called ``forkify'', for automatically converting unstructured loops with an affine induction variable and dynamic constant bounds into fork-joins. Fork-joins are easier to analyze than unstructured loops, including the degree to which they are parallel: all nodes dependent on the thread IDs of the fork and not in a cycle with a reduce node of the corresponding join can be executed in parallel. Forkify is implemented using standard loop dependence techniques.

\subsection{Optimizations and Parallelization}
\label{optimization}

We have implemented several transformations for optimizing Hercules IR. These include simple scalar optimizations, collection-oriented transformations, fork-join manipulation routines, and many others. This large set of optimizations enables users to transform their single application into device-specific optimized programs. Users can refer to parts of a Hercules IR module using the labels given in the source Juno code and passes in Hercules propagate labels and can be run on sub-graphs of functions, indicated by labels, allowing users to optimize specific portions of their code.

Users may apply \emph{attributes} to specific nodes indicating extra information about a node to the compiler. 
There are currently four attributes. \texttt{ParallelReduce} indicates that a particular reduction may occur in parallel, as the reduction consists of reads and writes to a collection that do not race. \texttt{MonoidReduce} indicates that a particular reduction may be re-associated, as the reduction computation is a monoid. \texttt{NoResetConstant} indicates that a particular collection constant doesn't need to be initialized to zeros, as it's fully initialized before it's ever read. \texttt{AsyncCall} indicates that a particular function call should be spawned in a parallel task. Attributes are often automatically inferrable, and when they are not the user can apply them in their schedule. Unlike transformation passes, manually applying attributes in schedules can be unsafe, as the compiler cannot verify that attributes are correct (for example, a misplaced \texttt{ParallelReduce} may cause data races).

A crucial utility to writing effective schedules  is a compiler debugger, as it is useful to see how transformations affect a program. Hercules includes a tool that visualizes a Hercules program as a Graphviz \cite{graphviz} plot at any point during a schedule---we found that a graphical view of a program is much easier to digest than a textual representation, since Hercules IR has the form of a semi-amorphous graph. Figures~\ref{fig:arr_sum_ir}~and~\ref{fig:matmul_ir} were generated using this tool.

\section{Device Code Generation and Composition}
\label{device_codegen}

Hercules IR is a target-agnostic program representation, meaning an application compiled with Hercules can be mapped onto heterogeneous hardware in many ways. We introduce a function outlining based approach to restructuring code for heterogeneous execution, which enables ``free'' generation of host orchestration code. We also describe a modified Global Code Motion algorithm which concretizes inter-device communication, and which eliminates most clones of collections.

\subsection{Device Partitioning via Outlining}
\label{outlining}

In Hercules, device backends operate on individual functions, but the hierarchy of functions in application code may not be the same as the hierarchy of device functions that should be emitted. For example, it may be profitable to fuse two functions, or it may be profitable to decompose a single reduction into a reduction tree, requiring multiple device kernels. To enable these transformations, we implement function inlining and outlining passes in Hercules. Inlining is implemented in a standard fashion. Outlining takes a subgraph of a function, indicated using labels, and replaces that subgraph with a call to a new function containing the outlined subgraph. Data flow edges crossing the subgraph's boundary either become parameters or return values of the callee, depending on their direction. These transformations allow developers to divide applications into functions based on application logic while using schedules to generate optimized code that uses a different structure of calls. We are not aware of another programming system that explicitly enables user-controlled, late-bound device partitioning.

The standard flow to perform device partitioning is to inline functions as needed to create large, flat functions, and then outline device functions. Framing device partitioning as function outlining has a nice side effect: host orchestration code is automatically generated. After outlining different subgraphs into device functions, the user is not left with several unrelated functions---the device functions are still called by a Hercules function whose body now consists mostly of function calls. This function correctly calls the device functions to produce an overall desired result and can orchestrate the execution of code across multiple devices. Thus, we frame host orchestration code generation as just another backend for a Hercules function, nominally for functions that call other functions (Section~\ref{rt_backend}). We note that this is not the norm for heterogeneous programming systems; prior systems, like HPVM \cite{hpvm2017} and Delite \cite{delite}, generate only device code and use a generic host orchestration runtime along with a bespoke description of how device functions should be invoked. After inlining and outlining, a mapped Hercules application can be viewed as an acyclic function call graph, where the internal nodes are host orchestration functions and the leaves are device functions.

We emphasize that since Hercules IR uses value semantics, any communication between the caller and callee automatically becomes explicit in the callee's signature. Flexible compile-time partitioning implies that inter-device communication is an \textit{emergent} property of the program, and thus it is essential that the compiler can automatically determine when it arises. In contrast, prior heterogeneous programming systems require explicit programmer annotations to identify inter-device communication prior to optimization. This greatly limits the flexibility of those prior systems.

\subsection{Global Code Motion}
\label{gcm}

After optimization and outlining, a Hercules IR module contains separate device and host functions. However, this code still uses value semantics, meaning that semantically every value, including collections, is created anew by every operation. To generate efficient code, we must convert the Hercules IR to use in-place mutation. This is performed as part of Global Code Motion (GCM), which assigns each collection to a device's memory and adds explicit copies, as needed, to allow for in-place mutation. Since in-place mutation only make sense when ordered, this is the phase at which we convert out of a sea-of-nodes representation and assign instructions into basic blocks.

Each control node becomes a basic block where the node's kind determines the branch at the end of the basic block. Data nodes are then placed into basic blocks according to their iterated uses and users using the schedule early/schedule late algorithm described by Click~\cite{gcm}. However, we modify this algorithm to avoid hoisting collection constants outside loops, which would otherwise cause unnecessary clones of the collection every loop iteration.

Next, we legalize in-place mutation of collections; to do this we rely on a simple observation: \textit{if a collection value is never implicitly cloned, all mutations can occur in-place}. Thus, the goal of this phase is to make implicit clones explicit. For example, if a collection value is used by two write nodes along the same control flow path, the second write induces an implicit clone of the original collection. We identify implicit clones and ``spill'' them by inserting a new collection constant and explicitly copying the cloned collection into it. In practice, we find that spilled clones are rare, at least in part because when assigning basic blocks, anti-dependencies between reads and writes are considered in order to place mutations after immutable uses when possible. Additionally, constant nodes which produce the initial value of a collection are placed inside loops, preventing that value from being live across loop iterations. We observed no spill clones when compiling any of the benchmarks in our evaluation suite.

We next statically determine which device each collection is allocated on. For device functions, each collection it uses must be on the device of the function. For host functions, the device of a collection depends on what device functions use that collection. We also identify collections that are needed on multiple devices, in which case we insert an explicit copy to a second collection to represent the necessary inter-device communication in the host code.

Since Hercules IR does not contain first class references and functions are not recursive, the size of allocations needed by a Hercules module can be determined statically (in terms of dynamic constants). Since dynamic constants are known upon calling a Hercules function, we can allocate all of the necessary memory upon entry and each generated host and device function will take a ``backing'' pointer argument that points to the memory for that function's collections.

\subsection{CPU Code Generation}
\label{cpu_backend}

Our CPU backend translates a Hercules IR function into a single threaded LLVM function---multi-core parallelism is exploited by the host code backend (Section~\ref{rt_backend}). Hercules IR constructs are simple to map into LLVM and since both Hercules IR and LLVM use SSA form, phis are lowered into LLVM directly. As GCM has already legalized the reference semantics of the Hercules IR function, operations on collections are lowered to loads and stores operating on pointers.

\subsection{GPU Code Generation}
\label{gpu_backend}

The GPU backend translates a Hercules IR function into a CUDA kernel and a thin host wrapper. The translation is mostly standard with sequential control flow translated directly into \texttt{gotos}. The key element of this backend is how we extract the parallelism present in fork-joins into the launch dimensions of the generated kernel. To do this, we assemble a tree of the fork-joins where fork-join $A$ is a child of fork-join $B$ if $A$ is contained within $B$ and $A$ is not contained in any of $B$'s other children, ensuring that this tree has a single root by adding a fork-join of factor 1 if there were originally multiple top-level forks. Then, traversing the tree in post order, we identify a valid launch size for each fork-join based on itself and its children. The root's launch size is then valid for the entire function. For each fork-join we calculate its launch size as follows:
\begin{enumerate}
    \item If the fork-join contains a sequential reduction, it must be sequential, and its launch size is the maximum size of its children (or 1 if childless).
    \item Otherwise, if the fork-join contains an associative reduction and the iterations will correspond to adjacent CUDA threads we can use CUDA's cooperative groups API. Therefore, we can either make the fork-join sequential or make its children sequential and set the launch size to its fork factor. We choose whichever option results in a larger launch.
    \item Finally, if every reduction is parallel then the launch size is the maximum size of its children, multiplied by its fork factor.
\end{enumerate}

The launch size determined by this procedure then needs to be split into blocks and threads. Hercules's policy is to spawn multiple blocks if the top fork-join contains only parallel reductions, and for all other fork-joins to enumerate threads in each block. Collection constants defined in a basic block inside a block-level fork-join are allocated in shared memory. Shared memory objects are given extra array dimensions if the collection constant node is also placed inside a thread enumerating fork-join. Associative reductions in leaf fork-joins are lowered into warp reductions.

\subsection{Host Code Generation}
\label{rt_backend}

The host code backend translates a Hercules IR function into an Asynchronous Rust~\cite{async-rust} function. This backend lowers code containing calls to other Hercules functions and is responsible for enabling multi-core parallelism through task-parallel function calls and data-parallel fork-joins. Multiple function calls that do not depend on each other can be run concurrently---if a function call is marked with the \texttt{AsyncCall} attribute, the host backend will launch that function call in an async closure and await it at the first use of the return value. Applying this attribute is always safe, but introduces the overhead of creating a concurrent task. Additionally, each iteration of fork-joins with only parallel reductions can be run in parallel, and the host backend lowers such a fork-join to an async closure containing the body, including the reduction. The host function launches a copy of this closure per thread and waits for all the threads to finish before allowing control flow to continue. Section~\ref{use_and_impact} shows an example of spawning chunks of a loop iteration space as parallel tasks. 
We use asynchronous Rust primitives to express multi-core parallelism because these primitives are independent of a specific runtime. Asynchronous Rust allows the programmer to pick a particular async runtime, from the many runtimes that exist (Tokio~\cite{tokio}, async-std~\cite{async-std}, etc), to execute their code.
These runtimes are widely used and high quality, and async Rust functions compose easily so it is straightforward for an application already using an async runtime to integrate code compiled with Hercules.

\subsection{The Host Rust Interface}
\label{rust_interface}

After generating device and host code for a Hercules application, we need a way to invoke it. For any Juno function marked with \texttt{\#[entry]}, we generate a Rust wrapper object called a \emph{runner}. A runner owns allocations of backing memory needed when calling the Hercules function, and since all allocations are determined statically, it only needs to make one allocation per device before invoking Hercules code.
Figure~\ref{fig:host_rust} shows an example of the host-side Rust code calling a Hercules function.
After instantiating a runner via the \texttt{runner!} macro, the Hercules function can be called with the \texttt{run} method. This method takes in two sets of arguments: dynamic constant arguments and data arguments. The dynamic constant arguments determine the sizes of allocated backing memories. Collection parameters and return values use custom reference types which incorporate Rust lifetimes to prevent memory errors, and implement convenience routines for conversion to and from Rust slices.

\begin{figure}
    \centering
\begin{lstlisting}[style=colouredRust]
use hercules_rt::{runner, HerculesRefInto};
juno_build::juno!("matmul");
async fn invoke_matmul(
    i: u64, j: u64, k: u64, a: &[f32], b: &[f32]
) -> Box<[f32]> {
    let mut r = runner!(matmul);
    let c = r.run(i, j, k, a.to(), b.to()).await;
    return c.as_slice().into();
}
\end{lstlisting}
    \caption{Rust calling a matrix multiply Hercules function.}
    \label{fig:host_rust}
\end{figure}

\section{Evaluation}
\label{eval}

We evaluate the Juno language and the Hercules compiler on 7 benchmark programs: a camera vision pipeline (CAVA)~\cite{hpvm2fpga.asap22}, an edge detection pipeline (EDGE)~\cite{hpvm2017}, and five benchmarks from the Rodinia benchmark suite~\cite{rodinia, rodinia2}: breadth-first search (BFS), speckle reducing antisotropic diffusion (SRAD), back-propagation (BP), and two computational fluid dynamics programs (EULER and PRE-EULER). We rewrote all of these benchmarks in the Juno application language and wrote separate schedules for targeting the CPU and GPU. For all experiments, we target a machine with an Intel Xeon Silver 4216 CPU with 16 cores and 32 threads, 192GB of memory, and an NVIDIA GeForce RTX 2080 Ti GPU with 11GB of memory. We measure wall clock time as a mean of 10 repeated executions on an otherwise idle system.

We took care to write the benchmarks' application logic in a straightforward way, since a goal of the Juno scheduling language is to optimize unmodified applications for many different machine configurations. Anecdotally, the time taken to write Juno schedules is mostly spent thinking of and profiling different optimizations. The language is expressive enough that large optimization decisions can be concisely expressed, often requiring changes to only single lines of code. This is different than past heterogeneous programming systems, where most optimizations have to be manually implemented in application code.

\begin{figure}
    \centering
    \begin{tikzpicture}
\begin{semilogyaxis}[
    height=0.73*\axisdefaultheight,
    width=\axisdefaultwidth,
    symbolic x coords={BFS, BP, EULER, PRE-EULER, SRAD, GEOMEAN},
    xtick=data,
    xticklabels={BFS, BP, EULER, PRE-\\EULER, SRAD, GEO-\\MEAN},
    x tick label style={font=\scriptsize, align=center},
    major tick length=0,
    minor x tick num={1},
    xminorgrids=true,
    log ticks with fixed point,
    ytick=\empty,
    extra y ticks={0.8, 1, 1.5, 2, 2.5, 3},
    extra y tick labels={0.8, 1, 1.5, 2, 2.5, 3},
    extra y tick style={major tick length=0.15cm},
    ylabel=Speedup over Hand-Written,
    ybar,
    legend style={at={(0.5,1)}, anchor=south, legend columns=-1},
    %
    point meta=exp(y),
    nodes near coords={\scriptsize\pgfmathprintnumber[fixed,fixed zerofill,precision=1]{\pgfplotspointmeta}},
    nodes near coords style = {
        small value/.style = { color=black, anchor=north, },
        large value/.style = { color=black, anchor=south, },
        placement/.code={%
            \begingroup
            \pgfkeys{/pgf/fpu}%
            \pgfmathparse{\pgfplotspointmeta<1.0}%
            \global\let\result=\pgfmathresult
            \endgroup
            \pgfmathfloatcreate{1}{1.0}{0}%
            \let\ONE=\pgfmathresult
            \ifx\result\ONE
                \pgfkeysalso{small value}%
            \else
                \pgfkeysalso{large value}%
            \fi
        },
        placement
    },
]
\addplot coordinates {(BFS,1.40) (BP,2.43) (EULER,0.86) (PRE-EULER,0.77) (SRAD,1.43) (GEOMEAN,1.25)};
\addplot coordinates {(BFS,0.91) (BP,2.83) (EULER,1.10) (PRE-EULER,0.87) (SRAD,2.83) (GEOMEAN,1.48)};
\draw[line width=0.2mm] (axis cs: {[normalized]4.5},.5) -- (axis cs: {[normalized]4.5}, 5);
\draw[line width=0.1mm] (axis cs: {[normalized]-0.5},1) -- (axis cs: {[normalized]5.5},1);
\legend{CPU,GPU}
\end{semilogyaxis}
\end{tikzpicture}
    \caption{Performance of benchmarks written in Juno and compiled with Hercules, compared against hand-written, device-specific baselines. The relative speedup of the Juno version is shown, higher is better.}
    \label{fig:hand-written-perf}
\end{figure}
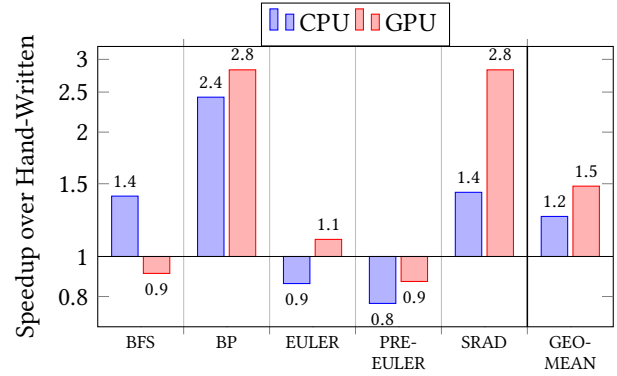

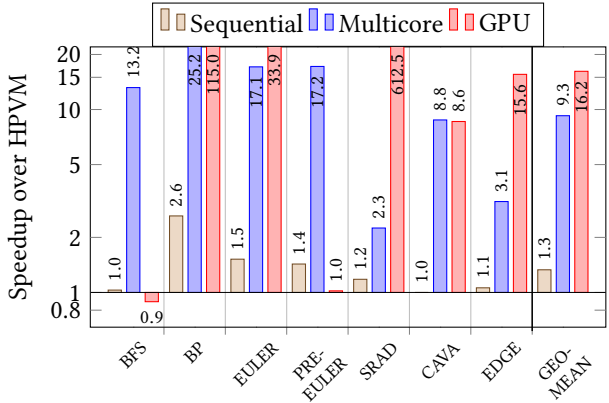
\begin{figure}
    \centering
    \begin{tikzpicture}
\begin{semilogyaxis}[
    height=0.73*\axisdefaultheight,
    width=\axisdefaultwidth,
    symbolic x coords={BFS, BP, EULER, PRE-EULER, SRAD, CAVA, EDGE, GEOMEAN},
    xtick=data,
    xticklabels={BFS, BP, EULER, PRE-\\EULER, SRAD, CAVA, EDGE, GEO-\\MEAN},
    x tick label style={font=\scriptsize, align=center, rotate=45},
    major tick length=0,
    minor x tick num={1},
    xminorgrids=true,
    log ticks with fixed point,
    ymax=22,
    ytick=\empty,
    extra y ticks={0.8, 1, 2, 5, 10, 15, 20},
    extra y tick labels={0.8, 1, 2, 5, 10, 15, 20},
    extra y tick style={major tick length=0.15cm},
    ylabel=Speedup over HPVM,
    ybar,
    bar width=5pt,
    legend style={at={(0.5,1)}, anchor=south, legend columns=-1},
    %
    point meta=rawy,
    nodes near coords={\scriptsize\pgfmathprintnumber[fixed,fixed zerofill,precision=1]{\pgfplotspointmeta}},
    log basis y=2,
    clip=false,
    restrict y to domain*=-1:4.5,
    nodes near coords style = {
        small value/.style = { color=black, anchor=north, rotate=0, },
        medium value/.style = { color=black, anchor=west, rotate=90, },
        large value/.style = { color=black, anchor=east, rotate=90, },
        placement/.code={%
            \begingroup
            \pgfkeys{/pgf/fpu}%
            \pgfmathparse{\pgfplotspointmeta<1.0}%
            \global\let\issmall=\pgfmathresult%
            \pgfmathparse{\pgfplotspointmeta>15.0}%
            \global\let\islarge=\pgfmathresult
            \endgroup
            \pgfmathfloatcreate{1}{1.0}{0}%
            \let\ONE=\pgfmathresult
            \ifx\issmall\ONE
                \pgfkeysalso{small value}%
            \else\ifx\islarge\ONE
                \pgfkeysalso{large value}%
            \else
                \pgfkeysalso{medium value}%
            \fi\fi
        },
        placement
    },
]
\pgfplotsset{cycle list shift=2}
\addplot coordinates {(BFS,1.03) (BP,2.62) (EULER,1.52) (PRE-EULER,1.43) (SRAD,1.18) (CAVA,1.00) (EDGE,1.06) (GEOMEAN,1.33)};
\pgfplotsset{cycle list shift=-1}
\addplot coordinates {(BFS,13.18) (BP,25.21) (EULER,17.11) (PRE-EULER,17.20) (SRAD,2.25) (CAVA,8.77) (EDGE,3.14) (GEOMEAN,9.25)};
\addplot coordinates {(BFS,0.89) (BP,115.01) (EULER,33.93) (PRE-EULER,1.02) (SRAD,612.48) (CAVA,8.60) (EDGE,15.57) (GEOMEAN,16.18)};
\draw[line width=0.2mm] (axis cs: {[normalized]6.5},.615) -- (axis cs: {[normalized]6.5}, 25);
\draw[line width=0.1mm] (axis cs: {[normalized]-0.7},1) -- (axis cs: {[normalized]7.7},1);
\legend{Sequential,Multicore,GPU}
\end{semilogyaxis}
\end{tikzpicture}
    \caption{Performance of benchmarks written in Juno and compiled with Hercules, compared against baselines written in Hetero-C++ and compiled with HPVM. The relative speedup of the Juno version is shown, higher is better.}
    \label{fig:hpvm-perf}
\end{figure}

\subsection{Performance}
\label{performance}

We compare Hercules's performance against prior hand-written implementations (using C/C++ and OpenMP on the CPU and CUDA on the GPU) of the Rodinia benchmarks~\cite{rodinia, rodinia2}, and against prior Hetero-C++ implementations compiled with HPVM~\cite{hpvm2017, hpvm2fpga.asap22}, a general purpose heterogeneous compiler, of all of the benchmarks. HPVM generates LLVM IR when targeting the CPU and OpenCL code when targeting the GPU. We compare both when mapping each application wholly onto the CPU and wholly onto the GPU.

Figure~\ref{fig:hand-written-perf} shows the relative speedup in wall clock time for each benchmark from the Rodinia suite on the CPU and GPU when compiled with Hercules, compared to the hand-written baseline programs. Hercules performs competitively with the aforementioned baseline codes on both the CPU and the GPU, achieving geomean speedups of $1.25\times$ and $1.48\times$, respectively. We observe considerable speedups on the CPU for the BP and SRAD benchmarks. For BP, this is due to the optimization of a particular subroutine---our schedule interchanges the two main loops, resulting in more efficient cache access. Additionally, we inline this subroutine at two call-sites that use different dimensions for arrays, allowing us to parallelize just the instance that operates on a large array. For SRAD, this is due to parallelizing a reduction that was not parallelized in the prior hand-written baseline. We see a slight performance degreation against the hand-written EULER and PRE-EULER; we believe this is due to their use of the \texttt{\#pragma omp simd} directive, which emits additional metadata instructing LLVM to perform aggressive vectorization. This causes LLVM's auto-vectorizer to behave differently between the hand-written baseline and Hercules generated code. With additional engineering effort, we could emit similar metadata when lowering fork-joins in the CPU backend. We also achieve considerable speedups on the GPU for the BP and SRAD benchmarks. A major bottleneck for both benchmarks are very large sum reductions and we use schedules to transform these loops into large, GPU only sum reduction trees using warp reduction instructions, while the baselines use sum reduction trees that are finished on the CPU and do not use warp reductions in the GPU portions. Additionally, in SRAD we fuse two large loops to eliminate a significant amount of global memory traffic. For BFS and PRE-EULER, we observe a slight degradation in performance which we attribute to the CUDA generated by Hercules containing more complex control flow, as Hercules IR represents unstructured control flow that is lowered directly into \texttt{gotos}.

Figure~\ref{fig:hpvm-perf} shows the relative speedup in wall clock time for each benchmark when compiled with Hercules, compared to when written in Hetero-C++ and compiled with HPVM. Like Hercules, HPVM can compile a target agnostic program onto either the CPU or the GPU. However, HPVM does not currently support generating multi-core code. To make a fair comparison, we include numbers for both sequential and multi-core execution of Juno programs. Hercules is performance competitive with HPVM when using a single core, and consistently faster when using multiple cores, achieving a $9.31\times$ geomean speedup. The largest sequential speedup is observed on BP; like the Rodinia baseline, the Hetero-C++ implementation of BP uses improperly ordered loops, resulting in poor cache access behavior. We copied this loop order in our Juno implementation, but use a schedule to interchange the loops. On the GPU, we see large speedups on all but one benchmark ($16.18\times$ geomean speedup). BP, SRAD, and EDGE all contain large reductions that HPVM lowers into slow sequential code on the GPU. HPVM IR does not properly represent generic parallel reductions, while Hercules IR does, and we can use simple schedules to assemble fast reduction trees on the GPU. We observe a slight slowdown on the GPU for BFS, which is for a similar reason as when comparing against the hand-written baseline: even though HPVM IR contains unstructured control flow, HPVM's OpenCL backend infers structured control flow. A similar analysis could be implemented in Hercules to generate structured control flow in CUDA with more engineering effort.

\begin{table}[]
    \centering
    \scriptsize
    \begin{tabularx}{\linewidth}{m{5.1em}|c|c|c|c|c|c|c}
    Transforms & BFS & BP & EULER & PRE-EULER & SRAD & CAVA & EDGE \\
    \hline
    Inlining & \cmark & \cmark & \cmark & \cmark & \cmark & \cmark & \xmark \\
    \hline
    Outlining & \cmark & \cmark & \cmark & \cmark & \cmark & \cmark & \cmark \\
    \hline
    Reduction\newline Transforms & \cmark & \cmark & \xmark & \xmark & \cmark & \xmark & \cmark \\
    \hline
    Loop Tiling \newline\& Reshaping & \cmark & \cmark & \cmark & \cmark & \cmark & \cmark & \cmark \\
    \hline
    Loop Fusion & \xmark & \xmark & \xmark & \xmark & \cmark & \cmark & \xmark \\
    \hline
    Loop Fission & \xmark & \cmark & \xmark & \xmark & \cmark & \xmark & \cmark \\
    \end{tabularx}
    \medskip
    \caption{What categories of transformations are used in the schedules for each benchmark.}
    \label{tab:opt_categories}
\end{table}

\begin{figure}
    \centering
\begin{lstlisting}[style=juno-scheduler]
macro reduction_tree![N](F) {
  fork-chunk![N](F);
  let (outer, inner) = fork-reshape[[0], [1]](F);
  monoid-reassociate(inner);
  let (top, bottom) = fork-fission(outer);
}
// In EDGE, max_gradient performs a large max reduction.
reduction_tree!(max_gradient);
\end{lstlisting}
    \caption{A macro to turn a sequential reduction into a reduction tree. \texttt{N} is the number of threads to spawn.}
    \label{fig:juno_reduction_tree_macro}
\end{figure}

\begin{figure}
    \centering
\begin{lstlisting}[style=juno-scheduler]
macro multicore![N](F) {
  fork-chunk![N](F);
  let (outer, inner) = fork-reshape[[0], [1]](F);
  let body = outline(inner);
}
\end{lstlisting}
    \caption{A macro to transform a parallel loop to exploit multi-core parallelism. \texttt{N} is the number of threads to spawn.}
    \label{fig:juno_multicore_macro}
\end{figure}

\subsection{Use and Impact of Schedules}
\label{use_and_impact}

Next, we elaborate on the contents of the Juno schedules for the benchmarks. Table~\ref{tab:opt_categories} summarizes the main categories of optimizations implemented in Hercules and indicates which optimizations we applied for each benchmark. We will also examine two schedules used for the benchmarks and their individual performance impacts more closely.

Almost every benchmark's schedule uses both inlining and outlining, as discussed previously these optimizations are used for restructuring applications into functions that map well onto specific devices. For example, to fuse two consecutive function calls, we first outline both callsites together into a new function, in which we inline the two calls in order to perform fusion. For many benchmarks, outlining is also used to convert one large function into several device functions. Outlining is also used to generate parallel loops targeting multi-core CPUs by outlining the body into a sequential CPU function leaving the loop structure in a host function, which can spawn loop iterations as asynchronous tasks.

The BP, SRAD, and EDGE benchmarks all contain large reductions that are converted into reduction trees. Hercules contains passes that optimize reductions---for example, when chunking a sum reduction, we can re-associate the sum such that each chunk sums starting at 0, and then the partial sums are added together at the very end, allowing the chunks to be run in parallel. These benchmarks also use fork fission to create multiple levels in the reduction tree---the first fork calculates the partial results in parallel, and the second fork serially combines them to produce a final result. This process can be repeated, for instance in the SRAD benchmark we further chunk and fission the second fork to create a reduction tree with 3 layers when targeting the GPU.

We use fork tiling, chunking, and reshaping in every benchmark, as they are needed to achieve good performance on both the CPU and the GPU. On the CPU, chunking partitions a loop into serial chunks that can be run in parallel. On the GPU, tiling enables generating multiple threads per block, which is necessary to achieve high warp utilization (where chunking places the newly created fork dimension before the divided dimension and tiling places it after).

Figure~\ref{fig:juno_reduction_tree_macro} shows a schedule for converting a sequential associative reduction into a reduction tree. The schedule is defined as a macro, which can be re-used on any loop containing an associative reduction. A reduction tree chunks up the input array, computes a partial reduction on each chunk, and then combines the partial reductions of each chunk. The schedule starts by chunking the loop (\texttt{fork-chunk!} and \texttt{fork-reshape}) and optimizing the reduction to start each partial reduction at the initial value of the reduction, rather than the previous partial result, so that partial results can be computed in parallel (\texttt{monoid-reassociate}). Finally, we convert the outer loop into two loops in a producer-consumer relationship (\texttt{fork-fission}). The first loop contains an inner loop computing partial reduction results, while the second loop contains the final reduction over the partial results. A multi-level reduction tree could be built by applying the \texttt{reduction\_tree!} macro to the produced bottom loop. We modified the GPU schedule for the EDGE benchmark to test its performance with and without constructing a reduction tree. We observe a $6.22\times$ end-to-end speedup from performing this transformation. Additionally, on the GPU we can map associative reductions of certain tile sizes to warp reduction primitives. We tile the per-thread sequential loop further by a constant factor, which allows the GPU backend to generate warp reduction primitives instead of sequential loops; with this we observe a total $9.77\times$ end-to-end speedup, compared to the sequential reduction loop.
Figure~\ref{fig:juno_multicore_macro} shows a schedule for converting a parallel loop into a form exploiting multi-core parallelism. Notice that Hercules uses the same small set of scheduling primitives (chunking, reshaping, outlining, etc.) both for parallelizing loops and for optimizing other parallel patterns, such as building reduction trees, demonstrating the power and versatility of the scheduling language. The schedule works by first chunking the loop by the number of threads that should be spawned (\texttt{fork-chunk!} and \texttt{fork-reshape}). It then outlines the inner loop into its own sequential CPU function leaving the outer loop in the caller function, which will be lowered by the host backend into parallel task launches. We modified the CPU schedules for every benchmark to test performance with and without chunking and outlining loop bodies in order to compare both parallel and sequential generated code against HPVM in Figure~\ref{fig:hpvm-perf}. We observe a geomean speedup of $7.0\times$ when using multiple cores, compared to sequential execution, across the Juno implementations of the benchmarks.

\section{Related Work}
\label{relwork}

\textbf{Low Level Device APIs}: Many device vendors ship low level programming systems designed for their hardware platforms, such as NVIDIA's CUDA~\cite{cuda_release} and PTX virtual ISA~\cite{ptx_release} and Xilinx's Vitis development suite~\cite{vitis}. These tools include a program representation to facilitate development, optimization, and mapping of device code onto vendor-specific hardware targets and a low level host API for interacting with the device, including facilities for initialization, task management, data movement, and synchronization. These systems are not productive to program for several reasons. First, they are vendor specific, meaning it is not feasible to write code using one vendor's tools that works on another vendor's device. Second, the programmer must manually implement high-level optimizations in their application code to achieve high performance, since the device programming models are often low level. Third, the programmer must manually write low level host code that controls the device---for sufficiently sophisticated applications, this host code is often highly coupled to the device logic, meaning both application and optimization changes require simultaneous device and host code edits. Some low level programming interfaces, such as OpenCL~\cite{tsuchiyama2011opencl}, HIP~\cite{hip_docs}, and Vulkan~\cite{vulkan_spec} eschew vendor specificity by working on multiple vendors' devices, but while this allows programmers to achieve functional portability, it often comes at the expense of performance, because different vendors' devices require different code transformations and vendor agnostic interfaces may not support advanced vendor-specific features.

\textbf{High Level Programming Models}: Several higher level programming systems have been proposed for productive programming of heterogeneous systems. Delite~\cite{delite} is a compiler for embedding high level DSLs inside Scala and compiling those DSLs onto heterogeneous machines. Like Hercules, Delite supports simultaneous execution of code on multiple hardware devices in a single machine. However, Delite's program representation is coarser than Hercules, as it contains coarse grained parallel patterns that are common across DSLs. Additionally, while Delite provides many transformations, code generators, and other components for flexible compilation, the example DSLs make their own decisions about how to optimize programs, and do not expose optimization control to the user. In contrast, Hercules gives the programmer explicit, granular control over how an application is lowered. We believe that exposing control over optimization is necessary for a \emph{general purpose} heterogeneous programming system, as the compiler cannot always infer how programs should be lowered onto different devices for arbitrary domains. Braid~\cite{braid} is a multi-stage programming system for heterogeneous programming. The authors identify placement and specialization as two key aspects of programming heterogeneous systems and propose static staging, rather than a scheduling language, to allow programmers to control for these factors. Additionally, Braid requires an in-application mechanism, called materialization, for safe inter-device communication, while in Hercules, inter-device communication is implicitly represented by data flow edges that cross an outlining boundary, requiring no changes to application code. Loo.py~\cite{loopy1, loopy2} is a code generator for array programs that is embedded inside Python---it exposes a library of program transformations and code generation debugging facilities as Python functions. Loo.py shares many of the same sensibilities as Hercules---for example, the provided kernel manipulation interface in Python is effectively an expressive imperative scheduling language. However, Loo.py's program representation is less general than Hercules, as it does not expose arbitrary control flow, making gradual discovery and exploitation of parallelism within larger programs infeasible. Additionally, Loo.py programs are individual kernels, rather than larger applications, requiring the user to perform manual application partitioning. HPVM~\cite{hpvm2017, hpvm2022} is a compiler infrastructure for general purpose heterogeneous programming. Like Hercules, HPVM uses a low level IR with constructs for parallelism and heterogeneity and has a general purpose frontend. HPVM, however, requires explicit annotations in application code to determine both parallelism and device partitioning. Additionally, HPVM does not feature as comprehensive a set of transformations, requiring the programmer to implement many device specific optimizations manually. Adding these transformations is not just a matter of engineering effort: some optimizations, like kernel fission, are particularly difficult to implement due to HPVM's memory model. None of these prior systems propose a general purpose and \textit{heterogeneous} scheduling language---the closest in prior work, Loo.py, offers a transformation library for intra-kernel optimizations, but does not address the problem of flexible application partitioning for heterogeneous systems.

\textbf{Domain Specific Systems}: Halide~\cite{halide} is a domain specific language and compiler for image processing programs. Halide uses a scheduling language to optimize image pipelines for many hardware targets and to compile image pipelines onto distributed memory systems~\cite{halide-dist}. While Halide also uses a powerful scheduling language, its scheduling language is declarative and specific to functional stencil computations in contrast to Hercules's imperative and general purpose scheduling language which can optimize general purpose imperative applications. Hercules's scheduler is more expressive and could serve as the basis for a library of both generic and domain-specific program optimizations. Several DSLs have been proposed for deep learning, such as TensorFlow~\cite{tensorflow2015-whitepaper}, PyTorch~\cite{pytorch, pytorch2}, and JAX~\cite{jax2018github} which can enable high performance portable code for heterogeneous systems by raising the level of programming abstraction. Many, but not all, of the program and runtime optimizations used in these systems are domain specific---Hercules could serve as a common program representation, optimization engine, and runtime executor for heterogeneous programs written using DSL frontends.

\section{Conclusion}
\label{conclusion}

This paper presents a programming system for productive programming of heterogeneous systems. We present Juno, a language for writing target agnostic application code and granular and expressive schedules. We also present Hercules, a compiler for extracting parallelism and heterogeneity in an application and targeting machines with multiple kinds of compute devices. Today, Hercules targets CPUs and GPUs, but it could be extended to support other architectures, such as FPGAs, in the future. Together, these components enable the separation of application development and performance optimization in a general purpose and heterogeneous context. We evaluate the viability of our approach by implementing the proposed system and demonstrating that it is capable of achieving high performance, both when compared against hand-written, device specific baselines and HPVM, a prior heterogeneous programming system.

\bibliographystyle{ACM-Reference-Format}
\bibliography{bib}
\end{document}